\documentclass[12pt]{article}
\usepackage[dvips]{graphicx}
\let\jnfont =\rm
\def\NPB#1,{{\jnfont Nucl.\ Phys.\ B }{\bf #1},}
\def\PLB#1,{{\jnfont Phys.\ Lett.\ B }{\bf #1},}
\def\PRD#1,{{\jnfont Phys.\ Rev.\ D } {\bf #1},}
\def\PRL#1,{{\jnfont Phys.\ Rev.\ Lett.\ }{\bf #1},}
\def\ZPC#1,{{\jnfont Z.~Phys.\ C }{\bf #1},}
\begin{document}
\begin{flushright}
AMES-HET-01-06 \\
April 2001 
\end{flushright}
\vspace{0.2in}
\begin{center}
{\Large Probing $R$-violating top quark decays at the NLC}
\vspace{.3in}

K.J. Abraham$^a$, Kerry Whisnant$^a$, Jin Min Yang$^b$, 
 Bing-Lin Young$^a$ 
\vspace{.3in}

{\small \it
$^a$ Department of Physics and Astronomy, Iowa State University,
     Ames, Iowa 50011, USA\\
$^b$ Institute of Theoretical Physics, Academia Sinica, Beijing 100080, China}
\end{center}
\vspace{.5in}
     \begin{center} ABSTRACT  \end{center}
We examine the possibility of observing exotic top quark decays
via $R$-Parity violating SUSY interactions in $e^{+}e^{-}$ collisions at 
$\sqrt{s}= 500~GeV$. 
We present cross-sections for $t\bar t$ production followed by the 
subsequent decay of either the $t$ or $\overline{t}$
via the $R$-Parity violating interaction while the other undergoes the 
SM decay. We discuss kinematic cuts that allow 
the exotic SUSY decays to be detected over standard model
backgrounds.
Discovery limits for $R$-Parity violating couplings in the top
sector are presented assuming an integrated luminosity of $100\,fb^{-1}$. 

\newpage
Even though there is no firm evidence contradicting the Standard Model,
it is widely believed to be the low energy approximation of a more 
fundamental theory. The currently most popular possible extension of the 
Standard Model is the minimal supersymmetric model (MSSM) which 
not only can account for all existing precision electroweak mesurements 
but also predicts a whole host of new particles and decays waiting 
to be observed. In its simplest form the MSSM has a global symmetry
$R$-parity \cite{rp}, 
defined by $R=(-1)^{2S+3B+L}$ with spin $S$, baryon-number $B$ and  
lepton-number $L$ which is implemented in such a way as to conserve both 
$B$ and $L$ seperately. However, there is no fundamental requirement for 
$R$ conservation, indeed it is known that 
instanton effects induce miniscule violations of both $B$ and 
$L$ \cite{tHooft}. Thus if SUSY is discovered and $R$-parity turns
out to be conserved, it is conceivable that some
hitherto unidentified fundamental principle is at work.
Hence $R$-parity violation should be vigorously searched for. 

Constraints on the $R$-parity violating couplings have been obtained 
from various analyses; however
as summarized in Ref.~\cite{review}, although many such couplings have been
severely constrained, existing bounds on the top quark couplings 
are generally quite weak. 
This is our motivation for the phenomenological study of R-violation in 
processes involving the top quark.
In what follows, we study the feasibility of detecting $B$ violating 
$R$-parity interactions ({\em i.e.} $\lambda^{''}$ couplings only)
in top production and decay in $e^{+}e^{-}$ collisions at $\sqrt{s}= 500~GeV$.
Given that
projected luminosities are large enough to produce tens of thousands of 
$\overline{t}t$ pairs in a relatively clean environment \cite{nlcrev},
it is reasonable to expect that either $R$-parity violating SUSY  
will be discovered or the parameter space further restricted.

Unlike the case of a hadron collider, where $B$ violating couplings lead to
new $\overline{t}t$ production mechanisms \cite{hadron}, 
at an $e^+e^-$ collider the effect of 
$B$ violating couplings has no effect on top pair production.  We thus focus 
on exotic top decay modes induced by $B$ violating couplings.
Furthermore,
we assume that the decay of either the $t$ or the $\overline{t}$ proceeds  
via the $R$-parity Violating interactions; one quark thus decays via 
Standard Model channels. With the restriction to $B$-Violating couplings
\footnote{For top decays induced by $L$ violating couplings, see 
\cite{tdecay_LV}.} only, the possible exotic decay modes are 
\begin{eqnarray}\label{BV}
t\to \tilde{\bar d_i}  \bar d_j,~\tilde{\bar{d}}_j \bar{d}_i
          \to \bar d_i  \bar d_j \tilde \chi^0_1 
\end{eqnarray} Sfermions involved in these decays can be on-shell or virtual,
depending on the masses of the particles involved. All calculations are 
performed using the narrow width approximation with top spin correllations 
taken fully into acount.

Among the decay modes which are relatively easy to detect are 
those induced by $\lambda^{''}_{3j3}$. Since $\lambda^{''}_{333}$ vanishes
due to the requirement of anti-symmetry on the final two indices we consider
only the cases where $i=1,2$. To keep the analysis simple we assume that 
either one, but not both, of the tri-linear coupling just mentioned takes a 
non-vanishing value. Our analysis can be easily extended to the case where 
both couplings are non-vanishing in the limit that the down type squarks are
approximately degenerate.

In our analysis we focus on the case of where $\lambda''_{313}$ is 
non-vanishing. As shown in Eq.(\ref{BV}), the decay 
$ t\to \bar b  \bar d \tilde \chi^0_1$ can proceed through exchange of
a sbottom ($\tilde b$)  or a down squark ($\tilde d$).  
Since among the down-type squarks the sbottom is most likely to be 
significantly lighter than others
\footnote{ There are arguments~\cite{heavysf} that first and second generation 
sfermions can be as heavy as 10 TeV without a
naturalness problem, while the third generation sfermions have
to be rather light.}, we assume the channel of exchanging 
a sbottom gives the dominant contribution.
Since only a light sbottom is meaningful to our analysis, the dominant 
decay mode of the sbottom is $\tilde b\to b \tilde \chi^0_1$. 
The charged current 
decay mode $\tilde b\to t \tilde \chi^+_1$ is kinematically forbidden for
a light sbottom in our analysis. We do not consider the strong decay mode
$\tilde b\to b \tilde g$ since the gluino $\tilde g$ is likely
to be heavy~\cite{CDF}. 

Note that the LSP ($\tilde \chi^{0}_{1}$) is no longer stable when R-parity is 
violated. In case just one R-violating top quark coupling does not vanish,
the lifetime of the LSP will be very long, depending on the coupling and the
masses of squarks involved in the LSP decay chain (cf. the last paper 
of~\cite{review}). We restrict ourselves to the same region of SUSY parameter 
space as in \cite{self}, leading to an LSP which decays outside the detector.

In this case, the final state in the exotic decay of the $t$ or $\overline{t}$
will consist of two jets
accompanied by missing energy. If in addition we consider purely hadronic
standard model $t(\overline{t})$ decays, we will have a very distinctive 
signal consisting of five jets and missing energy. 
(The inclusion of semi-leptonic
standard model decays not only does not increase the signal by much but 
also complicates the reconstruction of the top pair due to multiple sources
of missing energy.)

The main standard model backgrounds are:

\begin{itemize}

\item $W^{+}W^{-}Z$ production 
with the subsequent decay of 
$W^{+}W^{-}$ to five partons and $Z\rightarrow\overline{\nu}\nu$

\item $Z+5\rm{jets}$ production.

\end{itemize}
In order to isolate the signal we impose the following phase space cuts:
\begin{itemize}

\item Each jet must have an energy of at least 20 GeV

\item There must be a missing $\mbox{p}_{\mbox{t}}$ of at
least 20 GeV 

\item The invariant mass of at least one combination of three
jets must lie within 10 GeV of $m_{t}$, and the invariant mass of two of
the these three jets must lie within 5 GeV of $m_W$.

\item The invariant mass of the remaining two jets and the
invisible particles must lie within 10 GeV of $m_{t}$.

\item The absolute value of the cosine of the angle made by
any jet with the beam axis must be larger than 0.9.

\item For all jets we require the $y_{ij}$ be larger than
0.001 for all values of $i$ and $j$, where $y_{ij}$ is defined by
\begin{displaymath}
\frac{2 \mbox{min}(E_{i}^{2},E_{j}^{2})(1-\cos{\theta_{ij}})}{s}
\end{displaymath} where $i$ and $j$ denote jet indices and run from 1 to 5.

\end{itemize}

With the cuts listed above, $WWZ$ production gives a background of 
less than one event with a luminosity of 100$fb^{-1}$ and is thus 
small. This estimate is based on the narrow width approximation
\cite{fawzi}; the deviation between the narrow width approximation and 
a more refined treatment \cite{6ferm} is not large enough to affect 
this estimate. Nonetheless, to be conservative we estimate one background
event from $WWZ$ production.

Estimating the background from (Z +5 jet) production is more tricky due
to the huge number of different graphs involved. Furthermore, large NLO
QCD corrections may be expected in multi-parton final states
\cite{moretti}.  Rather than attempt an exact calculation (which would
be beyond the scope of this letter), we will use the numerical results
for 6 jet production \cite{moretti} to put an upper bound on this
background. The cross-section for 6 jet production at $\sqrt{s}$ = 500
GeV with the $y$ cut alone is 22 fb, adding the other cuts listed above
reduces the phase space by a factor of 200. We may thus use as an upper
limit on the cross-section for (Z + 5 jet) production a value of about
.15 fb, including a K--factor of 1.5 to be conservative.  Taking into
account the Z branching fraction to $\nu\bar\nu$, leads us to a
cross-section of .03 fb, corresponding to an irreducible background of 3
events with an integrated luminosity of 100 $\mbox{fb}^{-1}$. Note that
in the event that the LSP mass is sufficiently far removed from the Z
mass, a cut on the invariant mass of the invisible particles may further
reduce this background.  Combining the two backgrounds gives a total of
four events.

Before presenting the results of the signal, we briefly discuss the SUSY 
parameters involved. The most important SUSY parameters relevant to our 
analysis are the coupling $\lambda''_{313}$ and the sbottom mass because
the signal cross section 
is proportional to $|\lambda''_{313}|^2$ and decreasing with sbottom
mass.  We will vary these two parameters to see the limits for the signal to
be observable. Other SUSY parameters involved are the lightest neutralino 
mass and its coupling to sbottom, which are determined by the parameters
$M, M^{\prime},\mu$ and $\tan\beta$.
$M$ is the $SU(2)$ gaugino mass and $M^{\prime}$ is the hypercharge
$U(1)$ gaugino mass. 
$\mu$ is the Higgs mixing term ($\mu H_1 H_2$) in the superpotential.
$\tan\beta=v_2/v_1$ is the ratio of the vacuum expectation values of the 
two Higgs doublets. We work in the framework of the general MSSM, but 
assume grand unification of the gaugino masses, which gives the relation
 $M^{\prime}=\frac{5}{3}M\tan^2\theta_W\simeq 0.5 M$.
The LEP experiments disfavored small $\tan\beta$ values \cite{LEP_higgs}.
The SUSY explanation of the recently reported value of the muon anomalous 
magnetic moment also requires a large $\tan\beta$ and a 
positive $\mu$\cite{mu_SUSY}.
In our calculation we choose the following representative set of values:
\begin{eqnarray} \label{para}
M=150 {\rm ~GeV}, \mu=300 {\rm ~GeV}, \tan\beta=10. 
\end{eqnarray}
The chargino and neutralino masses in units of GeV are then given by
\begin{eqnarray}
& & m_{\tilde\chi^+_1}=133, ~m_{\tilde\chi^+_2}=328,~\nonumber\\
& & m_{\tilde\chi^0_1}=72,~m_{\tilde\chi^0_2}=134,
~m_{\tilde\chi^0_3}=308, ~m_{\tilde\chi^0_4}=327.
\end{eqnarray}

It should be remarked that SUSY parameters are generally not well-constrained
experimentally at the present time.  The only robust constraints are the
LEP and Tevatron lower bounds on some of the sparticle masses.
Therefore, the above SUSY parameter values used in our calculation are 
not the only choice.  They are a set of representative values which are 
allowed by the current experimental bounds.

Table~1 illustrates the dependence of the signal cross-section
on $m_{\tilde{b}}$ assuming $\lambda^{''}_{313}=1$. As can be seen, the 
signal drops sharply once the sbottom mass approaches $m_{t}$.
The values of $\lambda^{''}_{313}$ and $m_{\tilde{b}}$ corresponding to
the discovery level ($5\sigma$) are displayed in Fig.~1.  For
comparison, the results of Tevatron Run 2 (2 fb$^{-1}$), Run 3 (30
fb$^{-1}$) and LHC (100 fb$^{-1}$) are also presented, which are taken
from \cite{self}, but renewed by using the new SUSY parameter values.
The current upper bounds on $\lambda''_{313}$, obtained from $Z$ decays
at LEP I \cite{Zdecay}, are about 0.5 at $1\sigma$ level and 1.0 at
$2\sigma$ level for squark mass of 100 GeV. For heavier squarks, the
bounds get weaker because of the decoupling property of the MSSM. So one
sees from Fig.~1 that for $0.1 < \lambda^{''}_{313} < 1 $, the signal is
observable for a sbottom lighter than about 160 GeV.

\begin{table}
\begin{center}
\begin{tabular}{||l|l|l|l|l|l|l|l|l|l||} \hline
$\sigma$~(fb) & 252 & 231 & 189 & 136 & 83 & 37 & 6.9 & .02 & .01 \\
\cline{1-10}
$m_{\tilde{b}}$~(GeV) & 100 & 110 & 120 & 130 & 140 & 150 & 160 & 170 & 180 
\\
\cline{1-10}
\end{tabular}
\caption{Signal cross section versus bottom squark mass for
$\lambda''_{313}=1$.}
\end{center}
\end{table}

In case of nonbservation, the exclusion ($2\sigma$) limits can be
obtained, and are displayed in Figs.~2 and 3.  Comparing with the limits
of Tevatron Run 2, Run 3 and LHC, one sees that NLC exclusion limits are
the best.  As can be seen in Fig.3, it is possible to put limits on the
branching fraction of the SUSY decay of order a few percent for a
sbottom lighter than 160 GeV.

Note that the signal contains like--sign $b$ quarks, in contrast to the
background. In case of a positive signal, $b$ tagging will present
additional evidence for non-standard physics.  The results can also be
applied to the case of the presence of $\lambda''_{312}$ with sbottom
replaced by strange-squark.

To summarize, we have calculated the cross-section for $R$-parity violating
$t$ decays in $e^{+}e^{-}$ collisions at $\sqrt{s}=500$GeV. The 
standard model backgrounds can be minimized with suitable cuts leading
to discovery bounds about as stringent as at the LHC \cite{self}.

\section*{Acknowledgments}

This work was supported in part by DOE grant No. DE-FG02-94ER40817.

\newpage

\begin{figure}[htb]
\hspace*{-1cm}
\includegraphics[height=15cm,width=12cm,angle =-90]{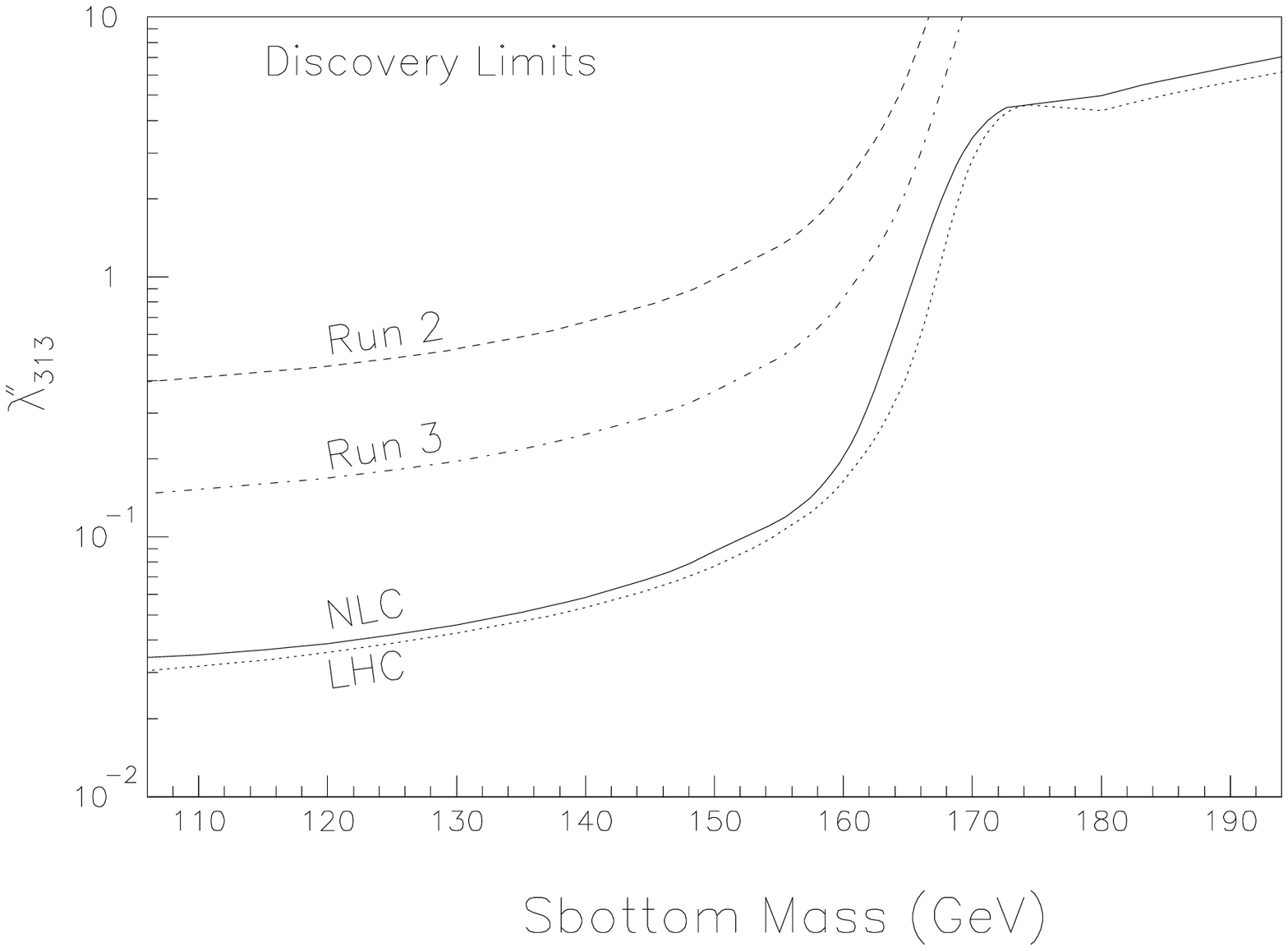}
\vspace*{-2cm}
\caption[]{ The discovery ($5\sigma$) limits of $\lambda''_{313}$ versus 
       sbottom mass. The region above each curve is the 
       corresponding region of discovery.  }
\label{fig.1}
\end{figure}
\newpage

\begin{figure}[htb]
\hspace*{-1cm}
\includegraphics[height=15cm,width=12cm,angle =-90]{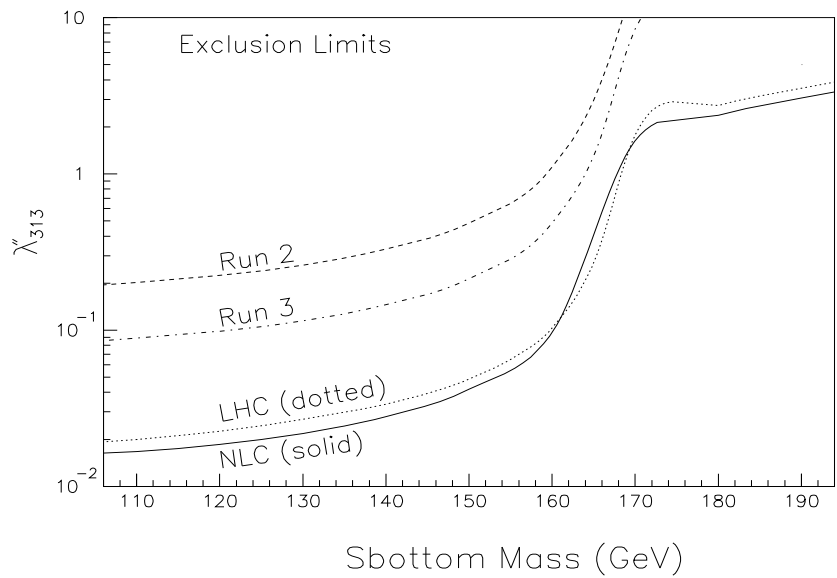}
\vspace*{-2cm}
\caption[]{ Same as Fig.~1, but for the exclusion ($2\sigma$) limits.
       The region above each curve is the corresponding region 
       of exclusion.  }
\label{fig.2}
\end{figure}

\newpage
\begin{figure}[htb]
\hspace*{-1cm}
\includegraphics[height=15cm,width=12cm,angle =-90]{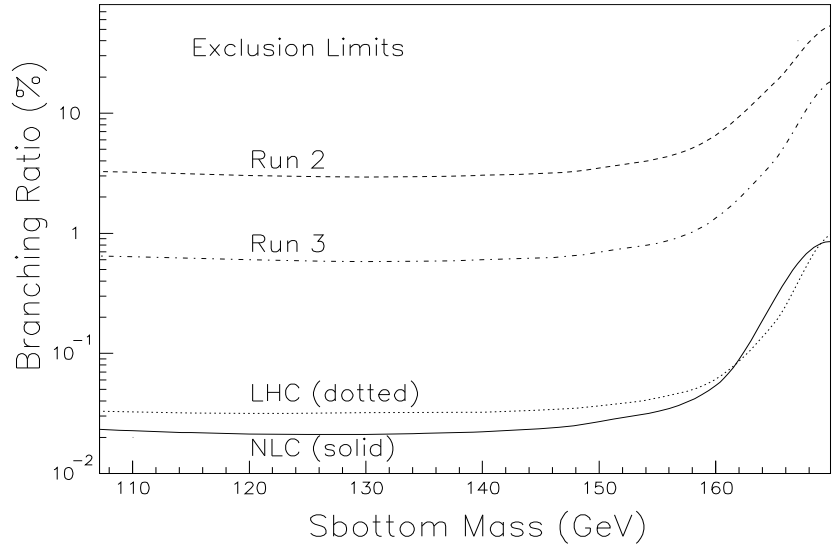}
\vspace*{-2cm}
\caption[]{ Same as Fig.~2, but for the branching ratio of the B-violating
       decay $t\to \tilde{\bar b}  \bar d$. }
\label{fig.3}
\end{figure}

\end{document}